\documentclass[12pt, draftclsnofoot, onecolumn]{IEEEtran}
\usepackage{amssymb}
\usepackage{graphicx}
\usepackage{amsmath}
\usepackage{epsf}
\usepackage{mathrsfs}
\usepackage{cite}

\newtheorem{corollary}{Corollary} 

\newtheorem{theorem}{Theorem}

\begin{document}
\title{{Strong Interference Alignment}}
\author{\Large Zainalabedin ~Samadi,  
       ~Vahid ~Tabatabavakili and ~Farzan ~Haddadi
\\\small Dept. of Elec. Eng.,   Iran University of Sceince and Technology 
Tehran,   Iran 
\\ \{z.samadi\}@elec.iust.ac.ir
\\ \{vakily,   haddadi\}@iust.ac.ir
}
\maketitle
\begin{abstract}

Interference alignment (IA) adjusts signaling scheme such that all interfering signals are squeezed in interference subspace. IA mostly achieves its performance via infinite extension of the channel, which is a major challenge for IA in practical systems. In this paper, we make part of interference very strong and achieve perfect IA within limited number of channel extensions. A single-hop $3$ user single antenna interference channel (IFC) is considered  and it is shown that only one of the interfering signal streams needs to be strong so that perfect IA is feasible.
\end{abstract}

\section{Introduction}

There are several strategies in multi user networks to manage interference. If interference is weak, the interfering signal is treated as noise. This approach has been used in practice for a long time, e.g., for frequency-reuse in cellular systems. However, information theoretic validation for this
approach has only recently been obtained \cite {Motahari09,   Shang09,   Annapur09}. On the other hand,  for the cases where  interference is strong, the interfering signal can be decoded along with the desired signal and hence canceled \cite{Carleial75,  Sato81,   Han81,   Sankar11,   Sridharan08}. However,  the general condition for strong interference in a $K>2$ user IC is unknown. The problem has been solved for some special cases such as symmetric IC. Lattice-based codes have been used to characterize a “very strong” regime  \cite{ Sridharan08},   the generalized degrees-of-freedom\cite{Jafar10},   and the approximate sum capacity  \cite{ Ordent12},   for symmetric $K$ user ICs.

There are cases where  strength of interference is comparable to the desired signal. Primary schemes,   such as time (frequency) division multiple access schemes,   avoid interference by orthogonally assigning the channel between users. Considering the entire bandwidth as a cake,   these schemes cut the cake equally between the  users. Therefore,   if there are $K$ users in the channel,   each user gets roughly $1/K$ of the channel. These orthogonal schemes,   however,   have been proved not to be bandwidth efficient. A recent strategy to deal with interference is IA. In a multiuser channel,   the IA method puts aside a fraction of the available dimension at each receiver for the interference and then adjusts the signaling scheme such that all interfering signals are squeezed in the interference subspace. The remaining dimensions are dedicated to communicate the desired signal, keeping it free from interference. IA was first introduced by Maddah Ali et. al. \cite{Maddah08},  and clarified in  \cite{JafarShamai}.

Cadambe and Jafar, \cite{Cadam08}, showed that the total number of $\eta= KM/2$  degrees of freedom (DoF)   can be achieved asymptotically via infinite time (frequency) expansion under block fading channel for a $K$-user $M$ antenna IFC. However, there is an important distinction between perfect IA schemes and partial IA schemes. Perfect IA schemes are able to exactly achieve the DoF outer bound with a finite symbol extension of the channel. In contrast, partial IA schemes pay a penalty in the form of the overflow room required to ”almost” align interference. 

Conventional IA schemes \cite{Cadam08} require global channel state information (CSI) including CSI of other communication links. Furthermore,  a huge number of dimensions based on time/frequency expansion are needed to achieve the optimal DoFs. These constraints need to be relaxed in order to apply IA to more practical systems.  Ergodic IA (EIA) scheme is proposed by Nazer et al.,  \cite{Nazer12} and seeks to send its signal in a favorable channel state condition. The order of channel extensions needed by  \cite{Nazer12} is roughly the same as \cite{Cadam08}.  EIA scheme  is based on a special pairing of the channel matrices and does not address the general structure of the paired channels  suitable for cancelling interference. \cite{Samadi} evaluates the general conditions on channel structure of an IFC to make  perfect IA possible with limited number of channel extensions.

 This paper addresses the  condition on channel matrices suitable for making part of interference very strong, so that this part of interference can be decoded and removed  prior to applying IA receiver filters. Assuming that part of interference is cancelled already, number of IA requirements reduces accordingly and perfect IA would be feasible in finite number of channel extensions.  In case the obtained structure on channel matrices does not meet a tolerable delay,   the method may turn into usual linear IA schemes.
  
  Notation: We use lower case for scalars,  boldface uppercase (lowercase) letters denote matrices (vectors).
  
\section{System Model} \label{secsysmod}

  \begin{figure}[h!] 
   \centering \includegraphics[scale=1.5]{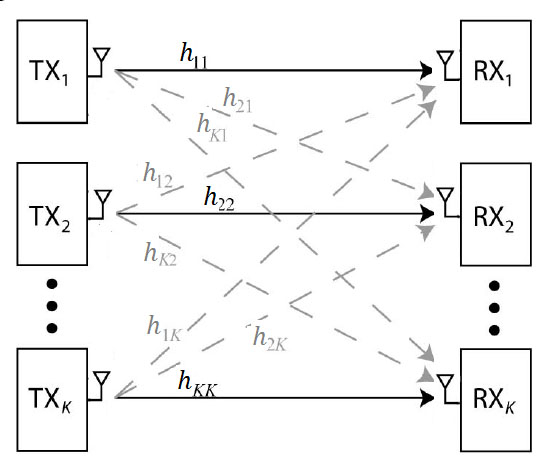}
  \caption{ Interference Channel Consisted of K Users, Each trying to Deliver its message to its intended receiver}
   \label{figure:KUser}
      \end{figure}

Consider a $K$ user single-hop ergodic single antenna interference network. An illustration of system model is shown  in Fig. \ref{figure:KUser}. Each transmitter wishes to communicate with its respective receiver. Communication takes place in a shared bandwidth and the goal is to achieve maximum possible  sum rate along with a reliable communication. 

The channel between transmitter $j$ and receiver $k$, $j, k\in\{1,  \ldots,  K\}$ at time instant $t\in \mathbb{N}$ is denoted as  $h^{[kj]} (t)$. It is assumed that values of channel coefficients at different time instants are independently drawn from a continuous distribution. The channel gains are bounded between a positive minimum value and a finite maximum value to avoid degenerate channel conditions.  The channel output observed by receiver  $k\in\{1,  \ldots,  K\}$  at time slot $t\in \mathbb{N}$ is a noisy linear combination of the inputs
\begin{eqnarray}
y^{[k]}(t)=h^{[k1]}(t)x^{[1]}(t)+h^{[k2]}(t)x^{[2]}(t) \cdots \nonumber \\+h^{[kK]}(t)x^{[K]}(t)+z^{[k]}(t), 
\end{eqnarray}
where $x^{[k]} (t)$  is the signal transmitted  by the $k^{\textrm{th}}$ transmitter,  and $z^{[k]}(t)$  is an additive independent and identically distributed noise drawn from a circularly
symmetric complex Gaussian distribution with unit variance, $z^{[k]}(t) \sim \mathcal{CN}(0, 1)$. It is assumed that all transmitters are subject to the power constraint:
\begin{eqnarray}
\mathrm{E}(\parallel x^{[k]}(t)\parallel ^{2}) \leq P, \quad 1 \leq k \leq K.
\end{eqnarray}
We assume that there is perfect channel state information (CSI) at receivers and transmitters. Hereafter,   time index is omitted for the sake of simplicity. 

Denote the capacity region of such a system with power constraint $P$ as $\mathcal{C}(P)$, corresponding DoF region is defined as
\begin{eqnarray} \begin{split}
\mathcal{D}= &\bigg \{ {\bf d}=(d^{[1]}, d^{[2]}, \ldots, d^{[K]}) \in \mathbb{R}_+^K: \\&\exists (R_1(P), R_2(P), \ldots, R_K(P)) \in \mathcal{C}(P), \\ & \textrm{such that}\; d^{[k]} = \lim_{P\rightarrow \infty} \frac{R_k(P)}{\log(P)}, \; k \in [{\bf K}] \bigg \}. 
\end{split}\end{eqnarray}

Total number of DoF is defined as $D_s= \max \sum_1^K d^{[k]}, \; \{d^{[1]}, d^{[2]}, \ldots, d^{[K]}\} \in \mathcal{D}$.

\section{Linear Vector IA Limitation}

Degrees-of-freedom region for a $K$ user IFC, with the system model discussed in section \ref{secsysmod}, has been derived in \cite{Cadam08} as follows, 
\begin{eqnarray}
\mathcal{D}= \left \{ {\bf d} \in \mathbb{R}_+^K: d_i+ d_j \leq 1, \; 1 \leq i, j\leq K \right \}, 
\label{dofreg}
\end{eqnarray}
and the number of DoF achievable by $K$ user IFC is obtained as $K/2$. It is straightforward to see that the following corollary describes the only DoF vector, ${\bf d}$, that achieves total number of DoF. 

\begin{corollary}
\label{cor1}
The only DoF vector that achieves   total number of DoF of an IFC  is
\begin{eqnarray}
d_i=\frac{1}{2}, \forall \; 1\leq i \leq K.
\label{optdof}
\end{eqnarray}
\end{corollary}

Consider a $3$ user IFC. We will use the scheme based on \cite{Cadam08} to perform IA. Assuming channel coefficients to be generic, it has been shown in \cite{Cadam08}  that optimal total number of DoF for a $3$ user IFC cannot be achieved over a limited number of channel usage. A brief review is presented here to maintain continuity of presentation. 

Let $\tau$ denote duration of the time expansion in number of symbols. Hereafter, we use the
upper case bold font to denote the time-expanded signals, e.g., ${\bf H}^{[jk]} = \textrm{diag} (h^{[jk]}(1) \ldots,h^{[jk]}(\tau))$, which is a $\tau \times \tau$ diagonal matrix. Denote the beamforming matrix of transmitter $k$ as ${\bf V}^{[k]}$. 

We intend to achieve the outer bound of  $3/2$ DoF for this setup. Consider $2n$ extension of the channel. Over this extended channel,  consider a hypothetical achievable scheme where each transmitter achieves $n$ DoF if possible,  using beamforming at every transmitter and zero-forcing at every receiver. Note that this is the only DoF point in achievable region that achieves total number of DoF of the network. The signal vector at  receiver $k$ can be stated as
\begin{eqnarray}
{\bf Y}^{[k]}&{}={}&{\bf H}^{[k1]}{\bf V}^{[1]}  {\bf X}^{[1] }+{\bf H}^{[k2]}{\bf V}^{[2]}  {\bf X}^{[2] }\nonumber \\ &&{+}\: {\bf H}^{[k3]}{\bf V}^{[3]}  {\bf X}^{[3] },+{\bf Z}^{[k]},
\end{eqnarray}
where ${\bf H}^{[k1]}$ is the $2n \times 2n$ extension of the channel, ${\bf V}^{[k]}$ is $2n \times n$   beamforming matrix of user $k$, and ${\bf X}^{[k]}$ is a $n\times 1$ column vector comprised of transmitted symbols  $x_m^{[k]},   m=1,  \ldots,   n $. An  encoder is assumed for user $k$ as $f_k : {0, 1}^{[2nR_k]} \rightarrow \mathcal{C}^n$, that map their respective inputs (the transmitted messages) to complex-valued sequences ${\bf X}^{[k]}$, each of block length $n$. Here $R_k$ denotes the bandwidth normalized
transmission rates (or spectral efficiencies) of user $k$. ${\bf Y}^{[k]}$ and ${\bf Z}^{[k]}$  represent the $2n$ symbol extension of $y^{[k]}$ and $z^{[k]}$,  respectively.

Receiver $i$ cancels the interference by  zero forcing all ${\bf V}^{[j]}, j\neq i$. The vectors corresponding to  interfering vectors must not occupy more than $n$ dimensions  from the $2n$ dimensional received signal vector ${\bf Y}^{[k]}$. Thus, IA requirements can be written as follows:

\begin{eqnarray} \begin{split}
\textrm{span} \left ( {\bf H}^{[ik]} {\bf V}^{[k]}  \right ) & =\textrm{span} \left ( {\bf H}^{[ij]} {\bf V}^{[j]} \right ), \\ & \quad i, j, k=1, 2, 3, \quad k, j \neq i, \end{split} 
\label{SE1}        
\end{eqnarray}
where $\textrm{span}({\bf A})$ denotes the column space of matrix ${\bf A}$. 

Note that the channel matrices $ {\bf H}^{[ji]}$ are full rank almost surely. Using this fact,   (\ref{SE1}) implies that 
\begin{eqnarray}
\textrm{span} \left (  {\bf V}^{[1]}  \right )=\textrm{span} \left ( {\bf T} {\bf V}^{[1]} \right ), 
\label{CAE1}           
\end{eqnarray}
where $ {\bf T}$ is defined as 
\begin{eqnarray}
 {\bf T}= ( {\bf H}^{[13]}  ) ^{-1} {\bf H}^{[23]}   ( {\bf H}^{[21]})^{-1} {\bf H}^{[12]} ( {\bf H}^{[32]}  ) ^{-1}  {\bf H}^{[31]}.
\label{TM}
\end{eqnarray} 

 (\ref{CAE1}) implies that at least one eigenvector of  ${\bf T}$ is in  $\textrm{span} \left (  {\bf V}^{[1]}  \right )$. Since all channel matrices are diagonal,   the set of eigenvectors for all channel matrices,   their inverse and product are all identical to the set of column vectors of the identity matrix,  namely vectors of the from ${\bf e}_k=[0 \;  \cdots \; 1 \; \cdots \; 0]^T$. Since ${\bf e}_k \in \textrm{span} \left (  {\bf V}^{[1]}  \right )$,   (\ref{SE1}) implies that 
\begin{eqnarray}
&{}& {\bf e}_k \in \textrm{span} \left ( {\bf H}^{[ij]} {\bf V}^{[j]}  \right ),   \quad \forall i,   j \in \{1,  2,  3\}.         
\label{imply}
\end{eqnarray}
Therefore,   at receiver $1$,   the desired signal $ {\bf H}^{[11]} {\bf V}^{[1]} $  is not linearly independent of the interference signal,   ${\bf H}^{[12]} {\bf V}^{[2]}$,   and  receiver $1$ cannot fully recover $ {\bf X}^{[1]}$ only by zero forcing the interference signal. Thus, $3/2$ degrees of freedom for the $3$ user single antenna IFC cannot be achieved through linear IA schemes, assuming channel coefficients to be completely random and generic.

\section{Strong Interference Alignment}
  Assume that one of the receivers (receiver $2$ for example) can linearly process its received $2n \times 1$ signal vector  to make part of the interference (e.g., signal of transmitter $3$) very strong. This strong interfering signal can be decoded and subtracted from the received signal, hence, one of the interfering beams is omitted at receiver $2$ and IA requirements can be ensured by perfectly aligning the remaining interferences as follows, 
\begin{eqnarray}
\rm{span} \left ( {\bf H}^{[12]} {\bf V}^{[2]}  \right )=\rm{span} \left ( {\bf H}^{[13]} {\bf V}^{[3]}  \right )
\label{SEp1}  \\ \rm{span} \left ( {\bf H}^{[23]} {\bf V}_{i^\prime}^{[3]}  \right ) \prec \rm{span} \left ( {\bf H}^{[21]} {\bf V}^{[1]}  \right ) 
\label{SEp2}        
\\ \rm{span} \left ( {\bf H}^{[31]} {\bf V}^{[1]}  \right )=\rm{span} \left ( {\bf H}^{[32]} {\bf V}^{[2]}  \right )
\label{SEp3}        
\end{eqnarray}
where ${\bf V}_{i^\prime}^{[3]}$  is the same as ${\bf V}^{[3]}$ except its $i^{\rm{th}}$ column being discarded because it is considered to be very strong.  ${\bf P} \prec {\bf Q} $ means that the set of column vectors of matrix ${\bf P}$ is a subset of the set of column vectors of matrix  ${\bf Q}$.

Therefore, we wish to pick vectors ${\bf V}^{[1]}, {\bf V}^{[2]}, {\bf V}_{i^\prime}^{[3]},  \rm{and}\; {\bf V}_{l}^{[3]}$ so that (\ref{SEp1}),
(\ref{SEp2}), (\ref{SEp3}) are satisfied along with very strong interference condition. Since the channel matrices are full rank; multiplying then by a full rank
matrix does not affect these conditions, (\ref{SEp1}),
(\ref{SEp2}), and (\ref{SEp3}), imply that 
\begin{eqnarray}
\rm{span} \left ({\bf T} ^{-1}{\bf V}_{i^\prime}^{[3]}  \right ) \prec \rm{span} \left ( {\bf V}^{[3]}  \right ) 
\label{SIAC}               
\end{eqnarray}
Consider the case of designing ${\bf V}^{[3]}$ as
 \begin{eqnarray}
{\bf V}^{[3]}  = \left [{\bf w}, {\bf T} {\bf w}, \ldots,  {\bf T}^{n-1}{\bf w} \right ]
\label{SEv3}               
\end{eqnarray}
in which, ${\bf w}$ is a $2n \times 1$ arbitrary column vector with no zero element. In this case, $i=1$ is the only choice that satisfies  (\ref{SIAC}). According to theorem \ref{theo1}, regardless of how ${\bf V}^{[3]}$ is designed,  the only one choice of  $i=i_0$ satisfies  (\ref{SIAC}).
\begin{theorem}
\label{theo1}
Assuming that ${\bf V}^{[3]}$ is a full rank $2n \times n$ matrix and  ${\bf T}$ is a random generic diagonal $2n \times 2n$ matrix, only one column of ${\bf V}^{[3]}$ can be chosen to satisfy (\ref{SIAC}). 
\end{theorem}
\begin{IEEEproof}

Suppose that, one can choose either one of ${\bf V}^{[3]}_{i_1}$ or ${\bf V}^{[3]}_{i_2}$ to satisfy (\ref{SIAC}). If the column $i_1$ of ${\bf V}^{[3]}$ is supposed to be a very strong interference,   (\ref{SIAC}) implies that ${\bf T} ^{-1}{\bf V}_{i_2}^{[3]} \in \rm{span} \left ( {\bf V}^{[3]}  \right )$. The same is true for ${\bf V}_{i_1}^{[3]}$  by considering ${\bf V}_{i_2}^{[3]}$ as the very strong interference stream direction. This implies that $\rm{span} \left ({\bf T} ^{-1}{\bf V}^{[3]}  \right )= \rm{span} \left ( {\bf V}^{[3]}  \right )$, which is the same as  (\ref{CAE1}) and therefore, IA will not be feasible. 
 \end{IEEEproof}
  
Assuming that ${\bf V}^{[3]}$ is designed as (\ref{SEv3}),   ${\bf V}^{[1]}$ and    $ {\bf V}^{[2]}$ can be obtained as follows, 
    \begin{eqnarray}
    {\bf V}^{[2]} =\left ({\bf H}^{[12]} \right)^{-1} {\bf H}^{[13]} {\bf V}^{[3]} 
\label{SEd2}  \\
 {\bf V}^{[1]} =\left ({\bf H}^{[31]} \right)^{-1} {\bf H}^{[32]} {\bf V}^{[2]}.
\label{SEd1} 
\end{eqnarray}
This set of transmitter beamforming matrices  $ {\bf V}^{[j]}, j=1,2,3$ satisfies IA requirements in   (\ref{SEp1}), (\ref{SEp2}),  and  (\ref{SEp3}). 

We wish to find a condition on channel coefficients such that
${\bf V}^{[3]}_{1} {\bf X}_{1}^{[3]}$ becomes very strong at receiver $2$. 

Assuming the linear processing of received signal vector at receiver $2$, the  interference resulting from  ${\bf X}_{1}^{[3]}$ becomes very strong if 
\begin{eqnarray}
C^{[23]}_1 \geq C^{[33]}_1, 
\label{vsic}
\end{eqnarray}
where $C^{[23]}_1$ is defined as 
 \begin{eqnarray} \begin{split}
C^{[23]}_1= & \mathrm{E}\big\{  \log \big ( \\ &1+P^{[3]}_1{{\bf V}^{[3]}_1}^H{{\bf H}^{[23]}}^H({\bf B}^{[23]})^{-1}){\bf H}^{[23]}{\bf V}^{[3]}_1\big ) \big \}.  \end{split}
\label{SIAR}
\end{eqnarray}
$P^{[3]}_1$ is transmitted power of stream $1$ from transmitter $3$, and ${\bf B}^{[23]}_1$ is the covariance matrix of noise plus interfering streams with respect to $ {\bf X}_{1}^{[3]}$. $C^{[33]}_1$ is defined as the achievable rate of ${\bf X}_{1}^{[3]}$ at its intended receiver, assuming that interference is totally cancelled at receiver $3$ by using IA. 

Suppose that the linear processing at receiver $2$ is performed as ${\bf c}_2^T {\bf y}_2$, where ${\bf c}_2$ is a $2n \times 1$ column vector to be designed in the sequel, and  ${\bf y}_2$ is the $2n \times 1$ received signal vector at receiver $2$. Received signal vector at receiver $2$ due to $ {\bf X}_{1}^{[3]}$ can be obtained as 
\begin{eqnarray}
I^{23}_1={\bf c}_2^H {\bf H}^{[23]}{\bf V}_1^{[3]}{\bf X}_1^{[3]}
\label{SIA1}
\end{eqnarray}
${\bf c}_2$ should be designed to maximize achievable rate of this signal at receiver $2$, which is, 
\begin{eqnarray}
C^{[23]}_1= \mathrm{E}\big\{ \log(1+\rm{SINR}^{[23]}_1) \big \}, 
\end{eqnarray}
where
 \begin{eqnarray}
\rm{SINR}^{[23]}_1=\frac{P^{[23]}_1}{1+P^{[21]}+P^{[22]}+P^{[23]}_{1^\prime}}.
\end{eqnarray}
$P^{[23]}_1$,  $P^{[23]}_{1^\prime}$ and  $P^{[2j]}, j=1, 2$ are defined as follows, 
 \begin{eqnarray}
 P^{[23]}_1={\bf c}_2^{H}{\bf H}^{[23]}{\bf V}^{[3]}_1  ({\bf V}^{[3]}_1)^H ({\bf H}^{[23]})^H {\bf c}_2 P^{[3]}_1\\
P^{[23]}_{1^\prime}={\bf c}_2^{H}{\bf H}^{[23]}{\bf V}^{[3]}_{1^\prime} {\bf P}^{[3]}_{1^\prime} ({\bf V}^{[3]}_{1^\prime})^H ({\bf H}^{[23]})^H {\bf c}_2
 \\
P^{[2j]}={\bf c}_2^{H}{\bf H}^{[2j]}{\bf V}^{[j]} {\bf P}^{[j]} ({\bf V}^{[j]})^H ({\bf H}^{[2j]})^H {\bf c}_2
\end{eqnarray}
where  ${\bf P}^{[3]}_{1^\prime}$ is the  diagonal matrix of transmitted powers for remaining streams other than stream $1$ from transmitter $3$,  and ${\bf P}^{[j]}, j=1,2$ is the diagonal matrix of transmitted powers from transmitters $j=1, 2$. 
Maximizing achievable rate  $C^{[23]}_1$ is equivalent to maximizing  $\rm{SINR}^{[23]}_1$. The optimum value for   ${\bf c}_2$ is evaluated as 
  \begin{eqnarray}
{\bf c}_2=\frac{({\bf B}^{[23]}_1)^{-1}{\bf H}^{[23]}{\bf V}^{[3]}_1}{\parallel({\bf B}^{[23]}_1)^{-1}{\bf H}^{[23]}{\bf V}^{[3]}_1\parallel}
\end{eqnarray}
where ${\bf B}^{[23]}_1$ is defined as 
\begin{eqnarray} \begin{split}
{\bf B}^{[23]}_1&=\sum_{k=1:3}{\bf H}^{[2k]}{\bf V}^{[k]}{\bf P}^{[k]}({\bf V}^{[k]})^H ({\bf H}^{[2k]})^H \\ &-{\bf H}^{[23]}{\bf V}^{[3]}_1  ({\bf V}^{[3]}_1)^H ({\bf H}^{[23]})^H P^{[23]}_1.  \end{split}
\end{eqnarray}
Using this value for ${\bf c}_2$,  $C^{[23]}_1$ is obtained as (\ref{SIAR}).

Therefore, the condition that interference due to  ${\bf X}_{1}^{[3]}$ becomes very strong is to have
\begin{eqnarray}
C^{[23]}_1 \geq C^{[33]}_1, 
\end{eqnarray}
where $C^{[33]}_1$ is obtained as follows, 
 \begin{eqnarray}\begin{split}
C^{[33]}_1&=\frac{1}{n}\mathrm{E}\big\{ \log( \\ & 1+{{\bf U}^{[3]}}^H{\bf V}^{[3]}{\bf H}^{[33]}{\bf P}^{[3]}{{\bf H}^{[33]}}^H{{\bf V}^{[3]}}^H{\bf U}^{[3]}) \big \},\end{split}
\end{eqnarray}
and ${\bf U}^{[3]}$ is   $2n \times n$ processing matrix that is designed to cancel aligned interference at receiver $3$. 

If (\ref{vsic}) is satisfied, ${\bf X}_{1}^{[3]}$  can be decoded first and subtracted from received signal, ${\bf Y}^{[2]}$, without compromising interference-free achievable rate of IA scheme. Receiver processing filters ${\bf U}^{[j]}, j=1, 2, 3$ can then be applied to the received signal streams, ${\bf Y}^{[1]}, {\bf Y}^{[2]}- {\bf V}^{[3]}_1{\bf X}^{[3]}_1, \rm{and} \; {\bf Y}^{[3]}$ to cancel the remaining interfering streams and extract desired signal vectors in a finite extension of the channel. 

In general, very strong interference condition (\ref{vsic}) can be considered for $l=1$ transmitted stream of transmitter $1$ as well. Therefore, channel aiding condition can be obtained as  
\begin{eqnarray}
C^{[2j]}_1 \geq C^{[jj]}_1, \quad j \in \{1,3\}. 
\label{GSIA}
\end{eqnarray}
If (\ref{GSIA}) is satisfied for a single $j \neq 2$, the very strong interference condition is satisfied and perfect IA is established. Note that strong interference is only required in one of the receivers (receiver $2$ in our case), interfering streams are simply cancelled in other receivers using zero forcing receiver matrices. Since we have assumed ergodic fading channel, expectation can be applied over time samples of the channel. 

\section{Simulation Results and Discussion}

 In this section we run simulations to illustrate practical feasibility of the strong interference assignment. It should be noted that, strong IA  can also be used jointly with linear IA schemes. User $2$ could ignore one of its desired signal streams instead of trying to make one of interfering streams very strong. In this case, the receiver zeroforcing filter for user $2$ is obtained as 
\begin{eqnarray}
{\bf U}^{[2]} _{m}= \nu_m[{\bf Q}^{[2]}], m=1, \cdots, n-1, 
\end{eqnarray}
where $\nu_m[{\bf A}]$ is the eigenvector corresponding to the $m^{\rm{th}}$ smallest eigenvalue of ${\bf A}]$. ${\bf Q}^{[2]}$ is the covariance matrix of interefering signals at receiver $2$ and is obtained as 
\begin{eqnarray}
{\bf Q}^{[2]} = \sum_{k=1, 3} \frac{P^{[k]}}{n} {\bf H}^{[2k]}{\bf V}^{[k]}{{\bf V}^{[k]}}^\dagger{{\bf H}^{[jk]}}^\dagger
\end{eqnarray}

 We can send $n_1$ symbols from  each transmitter,  using the precoding matrices ${\bf V}^{[1]},   {\bf V}^{[2]}$ and ${\bf V}^{[3]}$. If the channel aiding condition (\ref{GSIA}) is satisfied for some $2n  \leq 2n_1$,    $n$ degrees of freedom are achieved for each user. If the channel aiding condition is satisfied,   we have $n_1\geq n$ transmitted symbols for receiver $1$ in a $n$-dimensional subspace,   the same is true for receivers $2$ and $3$. We can resend $n_1-n$ symbols from each transmitter in consequent transmissions. Otherwise, if channel aiding condition is not satisfied for $2n \leq 2n_1$,  we proceed with the  conventional linear interference scheme to achieve $(n_1,  n_1-1,  n_1)$ degrees of freedom for the users in $2n_1$  channel uses. The steps of the strong IA are summerized in Table $1$.
  
  \begin{table} 
  \caption{ Strong IA Algorithm for $3$ User IFC.}
\centering 
\begin{tabular}{l}
\hline
1: One of the streams is regarded to be very strong in one of its unintended receivers. \\ 
2:  Assuming ${\bf X}_{1}^{[3]}$ is considerred to be very strong, compute  precoding matrices $ {\bf V}^{[j]}$\\ \nonumber \;  \; using \ref{SEv3} , \ref{SEd2}, and \ref{SEd1}.  \\
1: Evaluate strong interfernce condition  (\ref{vsic})  for  $2n<2n_1$, where $n_1$ is the number of \\ \nonumber \;  \; tranmitted symbols from each transmitter. \\ 
4: If (\ref{vsic}) is satisfied for any  $2n<2n_1$, ${\bf X}_{1}^{[3]}$  can be decoded first and subtracted from received \\ \nonumber \;  \;  signal, ${\bf Y}^{[2]}$. Receiver processing filters ${\bf U}^{[j]}, j=1, 2, 3$ can then be applied to the received  \\ \nonumber \;  \;  signal streams, ${\bf Y}^{[1]}, {\bf Y}^{[2]}- {\bf V}^{[3]}_1{\bf X}^{[3]}_1, \rm{and} \; {\bf Y}^{[3]}$. Go to step $6$.\\
5:  If  (\ref{vsic}) is not satisfied, proceed with the  conventional linear interference scheme to achieve \\ \nonumber \;  \; $(n_1,  n_1-1,  n_1)$ DoF for the users in $2n_1$  channel uses. \\
6: Continue till all messages are sent. \\
\hline
\end{tabular}
\end{table}

 \begin{figure}[h!]
 \centering \includegraphics[scale=0.65]{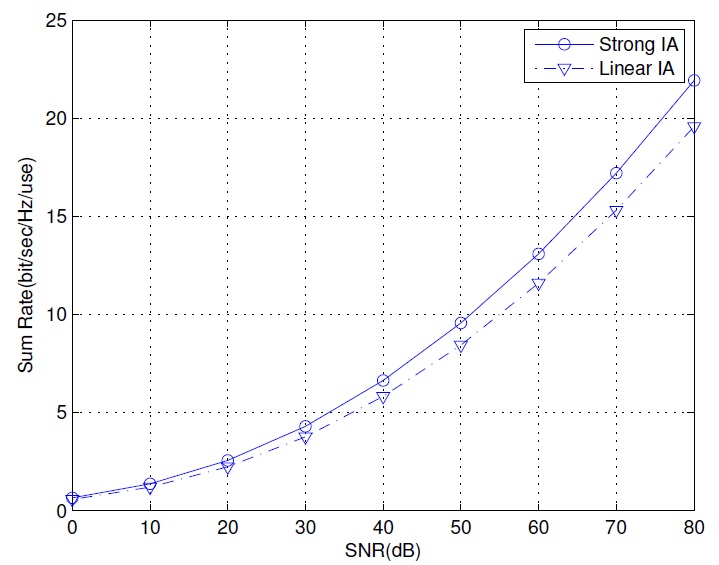}
  \caption{ Achievable Sum Rate Using Strong IA and Linear IA. Improvement is visible using strong IA versus  conventional linear IA.}
   \label{figure:StrongIA}
      \end{figure}
 
Consider a $3 \times 3$ user IFC. $6$ and $7$ extensions of the channel is used for strong IA and linear IA developed in \cite{Cadam08}, respectively. Sum rate performance of Strong IA is compared to that of Linear IA in Fig. \ref{figure:StrongIA}.  Channel coefficient are assumed to be i.i.d Gaussian random variables. Strong IA advantage over Linear IA schemes could be more substantial if number of users are more than 3 and/or some of the interfering links are inherently strong. However, the exact extent of improvement needs elaborate analysis and simulations over the specific scenario. 
 
 \section {Conclusion}
In this paper,   we proposed a new scheme called strong interference alignment  to achieve optimal degrees of freedom of the IFC using finite extension of the channel. This approach cancels part of the interference that is strong to ease IA requirements. The proposed method can also be used jointly with linear IA schemes. In case the channel coefficients do not meet the strong interference requirement obtained in (\ref{GSIA}), usual linear IA can still be used without needing to reconfigure the transmit precoding matrices. In total, strong IA method aims to reduce the required dimensionality and signal to noise ratio for exploiting degrees of freedom benefits of IA schemes.


\begin{thebibliography}{9}

\bibitem{Motahari09} 
A.~S. Motahari,   and A.~K. Khandani,   'Capacity bounds for the gaussian interference channel,    \emph{IEEE Trans. Inform. Theory},   2009,   \textbf{55}, (2),   pp. 620--643.

\bibitem{Shang09} 
X.~Shang,   G.~Kramer and B.~Chen,   'A new outer bound and the noisy-interference sum rate capacity for gaussian interference channels,   
  \emph{IEEE Trans. Inform. Theory}, 2009,   \textbf{55}, (2),   pp. 689--699.

\bibitem{Annapur09} V.~S. Annapureddy,   and  V.~V. Veeravalli,   'Gaussian interference networks: sum capacity in the low-interference regime and new outer bounds on the capacity region,    \emph{IEEE Trans. Inform. Theory},  2009,   \textbf{55}, (7),   pp. 3032--3050.


\bibitem{Carleial75} 
A.~B. Carleial,   'A case where interference does not reduce capacity,   
  \emph{IEEE Trans. Inform. Theory}, 1975,   \textbf{IT-21},   pp. 569--570.

\bibitem{Han81}
T.~S. Han and K.~Kobayashi,   'A new achievable rate region for the interference  channel,    \emph{IEEE Trans. Inform. Theory},   \textbf{IT-27},  Jan. 1981,   pp. 49--6o.

\bibitem{Sankar11} 
L. ~Sankar,    Xiaohu Shang,   E.~ Erkip,   and H.~V. Poor,   'Ergodic fading interference channels: sum-capacity and separability,    \emph{IEEE Trans. Inform. Theory}, 2011,   \textbf{57}, (7),   pp. 2605--2626.

\bibitem{Sridharan08}
 S.~Sridharan,    A.~Jafarian,   S.~Vishwanath,   and S.~A. Jafar,   'Capacity of symmetric K-user gaussian very strong interference channels,    in \emph{Proc. IEEE Global Commun. Conf.},   New Orleans,   LA,   Dec. 2008,   pp. 1--5.
 
 \bibitem{Jafar10} 
S.~ A. Jafar,  and S.~Vishwanath,   'Generalized degrees of freedom of the symmetric gaussian K user interference channel,    \emph{IEEE Trans. Inform. Theory}, 2010,   \textbf{56}, (7),   pp. 3297--3303.

\bibitem{Ordent12} 
O.~Ordentlich,   U.~ Erez,   and B.~Nazer,    'The approximate sum capacity of the symmetric gaussian K-user interference channel,    \emph{IEEE Trans. Inform. Theory}, 2014,   \textbf{60}, (6),   pp. 3450--3482.
  

\bibitem{Maddah08} 
 M.~ A. Maddah-Ali,   A.~S. Motahari,    and A.~K. Khandani,    'Communication over MIMO X channels',  2008,    \textbf{ 54},    (8),    pp. 3457--3470.
 
 \bibitem{JafarShamai} 
 Jafar, S. A.  and Shamai (Shitz), S.,    'Degrees of freedom region for
the MIMO X channel, \emph{IEEE Trans. Inform. Theory},  2008,     \textbf{ 54},    (1),    pp. 151--170.
 

\bibitem{Cadam08} 
 V.~ R. Cadambe,    and S.~ A. Jafar,    'Interference alignment and degrees of freedom of the  K-user interference channel,    \emph{IEEE Trans. Inform. Theory},  2008,    \textbf{ 54},    (8),    pp. 3425--3441.
 
 \bibitem{Jung11} 
Jung,  B.~C.,  Shin,  W.~Y.: `Opportunistic interference alignment for interference-limited cellular TDD uplink',   \emph{IEEE Commun. Lett.},  2011,  \textbf{15},  (2),  pp. 148--150.

\bibitem{Yang13} 
Yang,  H.~J., Shin,   W.~Y.,  Jung,   B.~C.,  Paulraj,  A.: `Opportunistic interference alignment for MIMO interfering multiple-access channels',   \emph{ IEEE Trans.  Wireless Commun.},  2013,  \textbf{12},  (5),  pp. 2180--2192.

\bibitem{Nazer12} 
Nazer, B.,  Gastpar,    M.,  Jafar,   S.~A.,  Vishwanath,   S.: 'Ergodic interference alignment',   \emph{IEEE Trans. Inf. Theory},  2012,   \textbf{58},   (10),    pp. 6355--6371.

\bibitem{Samadi}
Samadi,   Z.,  Vakili,    V.~T.,  Haddadi,   F.: 'Channel Aided Interference Alignment',   \emph{IET Signal Processing},  2017,   \textbf{11},   (7),    pp. 854--860.

\end{thebibliography}
\end{document}